\documentclass[a4paper,draft]{book}
\usepackage{epsf,nano2cmr,amssymb}
\begin{document}

\pnum{} \ttitle{Quantum derivatives and high-frequency gain in
semiconductor superlattices}

\tauthor{{\em A.~V.~Shorokhov} and K.~N.~Alekseev}

\ptitle{Quantum derivatives and high-frequency gain in
semiconductor superlattices}

\pauthor{{\em A.~V.~Shorokhov}$^{1}$ and K.~N.~Alekseev$^{2}$}

\affil{$^{1}$~Institute of Physics and Chemistry, Mordovian State University, 430000 Saransk, Russia\\
$^{2}$~Department of Physical Science, P.O. Box 3000, University
of Oulu FI-90014, Finland}

\begin{abstract}
{We derive simple difference formulas describing a small-signal
absorption of THz field in a semiconductor superlattice driven by
a microwave pump. We give a transparent geometric interpretation
of these formulas that allows a search of optimum conditions for
the gain employing only a simple qualitative analysis. Our
theoretical approach provides a powerful tool for finding the
correspondence between quasistatic and dynamic regimes in
ac-driven semiconductor superlattices.}
\end{abstract}

\begindc

\index{Shorokhov A. V.}
\index{Alekseev K. N.}

\section*{Introduction}
Semiconductor superlattices (SSL) have attracted growing attention
in view of their unique electronic properties, which can be used
for generation, amplification and detection of a high-frequency
electromagnetic radiation \cite{wackerrew}. Nonlinear transport
properties of SSL allows a generation of THz radiation in
conditions of negative differential conductance (NDC)
\cite{Kti72}. However, the static NDC makes SSL unstable against
formation of high-field electric domains \cite{Ign87}. These
electric domains are believed to be destructive for the THz gain
in SSLs. Currently the main focus is on the possibilities to
overcome this drawback within the scheme of dc-biased SSL
\cite{Sav04,Kre11}. However, schemes of THz superlatice devices
with ac pump fields are also under discussion
\cite{Ale05_2,Klap04}. A strong microwave field
$E_p(t)=E_1\cos(\omega_1t)$ pumps the SSL and a desirable signal
field $E_s(t)=E_2\cos(\omega_2t)$ has a higher frequency
$\omega_2>\omega_1$. Because for typical SSLs the characteristic
scattering time $\tau$ at room temperature is of the order of
$100$ fs, an interaction of the microwave fields with the miniband
electrons is quasistatic ($\omega_1\tau\ll 1$). Importantly, we
have showed recently that such quasistatic pump field can
completely suppress domains in SSLs \cite{Ale05_2}. Two distinct
possibilities exist for the signal field: It can be also
quasistatic ($\omega_2\tau\ll 1$) or it cannot be described within
the quasistatic approach if $\omega_2\tau\gtrsim 1$. The later
situation, i.e. $\omega_1\tau\ll 1$ but $\omega_2\tau\gtrsim1$,
can be called \textit{semiquasistatic interaction}. Thus, the
semiquasistatic approach is introduced to describe an
amplification of THz field in SSL under the action of microwave
field.
\begin{figure}[b]
\leavevmode \centering{\epsfbox{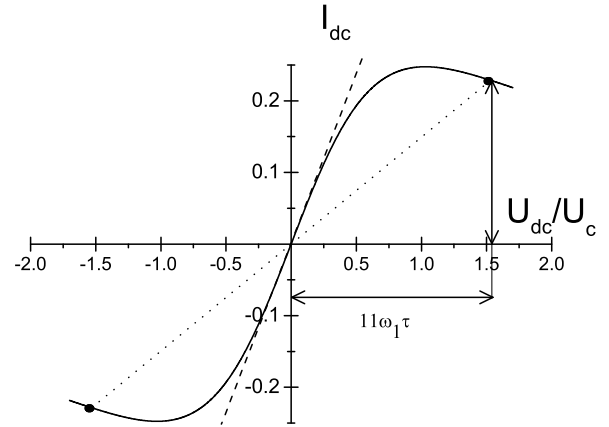}} \caption{\label{fig:1}
Geometrical meaning of the incoherent component of absorption
within semiquasistatic and quasistatic approaches. Time-averaged
current $I_{dc}$ under the action of quasistatic pump
($\omega_1\tau=0.1$, $U_{ac}/U_c=0.2$) vs dc voltage $U_{dc}$. If
we choose working point at $U_{dc}=0$, then the dotted segment
corresponds to the finite difference for the weak probe at
$\omega_2=11\omega_1$, the dashed straight line corresponds to the
derivative. The slopes of these strait lines determines
$A^{incoh}$ in the semiquasistatic and quasistatic cases,
respectively.}
\end{figure}
\begin{figure}[b]
\leavevmode \centering{\epsfbox{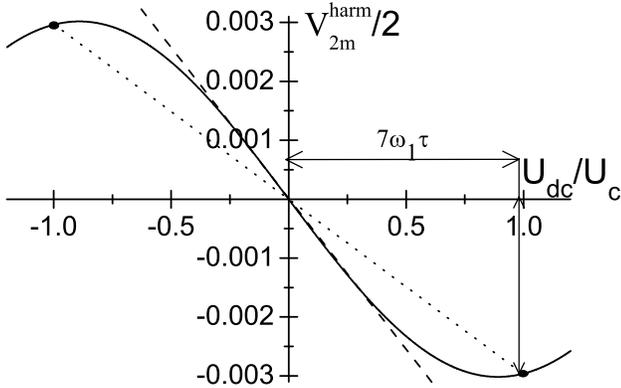}} \caption{\label{fig:2}
Geometrical meaning of the coherent component of absorption witin
semiquasistatic and quasistatic approaches. The $2m$th harmonic of
quasistatic current $V_{2m}^{harm}$ vs dc voltage $U_{dc}$ for
$m=7$. The current is induced by the quasistatic pump
($\omega_1\tau=0.14$) with the amplitude $U_{ac}/U_c=8$. If we
choose working point at $U_{dc}=0$, then the dotted segment
determines to the finite difference for the weak probe at
$\omega_2=7\omega_1$, the dashed stright line corresponds to the
derivative. The slopes of these lines determines $A^{coh}$ within
the semiquasistatic and quasistatic cases respectively.}
\end{figure}

\section{Quantum derivatives}
Our main findings are following. We consider the response of
miniband electrons to an action of the total electric field $
E(t)=E_0+E_1\cos(\omega_1t)+E_2\cos(\omega_2t)$, where $E_0$ is
the dc bias and  $E_{s}=E_2\cos(\omega_2t)$ ($\omega_2=m\omega_1$,
$m\in\mathbb{N}$) is the weak signal (probe) field. In the real
devices $E_{s}$ is a mode of the resonator tuned to a desirable
THz frequency.

We define the dimensionless absorption of the weak ac probe field
in SSL as $A(\omega_2)=2\langle V(t)\cos(\omega_2 t)\rangle_t$,
where $V(t)$ is the miniband electron velocity defined in the
units of the maximal miniband velocity, averaging
$\langle\ldots\rangle_t$ is performed over time. Starting from the
exact formal solution of the Boltzmann transport equation, we
represent the absorption $A$ as a sum of three terms
 \cite{Ale05_2,Pav77}
\begin{equation}
A=A^{harm}+A^{coh}+A^{incoh}.
\end{equation}
Here $A^{harm}$, $A^{coh}$ and $A^{incoh}$ describe the absorption
(gain) seeded by generation of harmonics, the parametric
amplification of the probe field due to a coherent interaction of
the pump and the probe fields, and the nonparametric absorption,
correspondingly. The derivation of semiquasistatic formulas is
based on the use of the asymptotic saddle-point method.
\par
Using this method we have get the following results. The term
$A^{harm}$ is just the expression for $m$th in-phase  harmonic of
the time-dependent current through SSL
\begin{equation}
\label{harm-gen} A^{harm}=2\left\langle
I(U_{dc}+U_{ac}\cos(\omega_1t))\cos(m\omega_1t)\right\rangle_t,
\end{equation}
where $U_{dc}=eLE_0$ is the dc voltage, $L=Nd$ is the length of
SSL ($d$ is the period of SSL and $N$ is the number spatial
periods), $U_{ac}=eLE_1$ is the amplitude of ac voltage created by
the pump field across SSL, the current $I(t)$ is normalized to the
maximal current in SSL, $I_0$,  corresponding to the maximal
miniband velocity. Note that $A^{harm}$ gives the main
contribution to the absorption of a weak probe \cite{Ale05_2}. The
expression for $A^{harm}$ is not specific for the semiquasistic
limit because of its independence on $\omega_2$. However, {\it our
main finding is that the absorption components $A^{coh}$ and
$A^{incoh}$ can be represented within semiquasistatic approach
using the specific terms of quantum derivatives as}
\begin{eqnarray}
\label{coh}
A^{coh}=e U_2\left\langle\frac{I^{ET}\big(U_{dc}+
U_{ac}\cos(\omega_1t)+N\hbar\omega_2/e\big)}{2N\hbar\omega_2}\right.\nonumber\\
\left.-\frac{I^{ET}\big(U_{dc}+
U_{ac}\cos(\omega_1t)-N\hbar\omega_2/e\big)}{2N\hbar\omega_2}\cos(2m\omega_1t)\right\rangle_t,\\
\end{eqnarray}
\begin{eqnarray}
\label{inc} A^{incoh}=e U_2\left\langle\frac{I^{ET}\big(U_{dc}+
U_{ac}\cos(\omega_1t)+N\hbar\omega_2/e\big)}{2N\hbar\omega_2}\right.\nonumber\\
\left.\frac{-I^{ET}\big(U_{dc}+
U_{ac}\cos(\omega_1t)-N\hbar\omega_2/e\big)}{2N\hbar\omega_2}\right\rangle_t,
\end{eqnarray}
where $U_2$ is the amplitude of small-signal voltage and
\begin{equation}
\label{volt} I^{ET}(U)=\frac{U/U_c}{1+(U/U_c)^2}
\end{equation}
is the Esaki-Tsu voltage-current (UI) characteristic ($U_c=\hbar
N/e\tau$ is the critical voltage and $I^{ET}$ is normalized to the
maximal current $I_0\equiv 2I^{ET}(U=U_c)$).
\par
Importantly, following Eq.~(\ref{inc}) in order to find the
incoherent absorption in SSL at arbitrary high frequency
$\omega_2$ we need to know only
\begin{equation}
\label{VI} I_{dc}(U_{dc})=\left\langle
I^{ET}\big(U_{dc}+U_{ac}\cos(\omega_1t)\big)\right\rangle_t,
\end{equation}
that is, \textit{the time-averaged current induced by the
quasistatic field (voltage)}. For a given amplitude of ac voltage
$U_{ac}$, the dc current $I_{dc}$ is a function of only dc bias
$U_{dc}$. It is easy to calculate or to measure the modifications
of UI characteristic caused by the action of microwave
(quasistatic) field \cite{Klap04a,Win97}. Note that in the
quasistatic case $A^{incoh}$ goes into a usual derivative of
current. The difference between these derivatives is shown on the
Figure~\ref{fig:1} where the dashed line is a derivative of dc
current and the dotted line is a quantum derivative depended on
the number of harmonic $m$. The importance of finding of
$A^{incoh}$ is stipulated by the following fact: $A^{incoh}$ plays
an essential role in the stabilization of space-charge instability
in SSL. We need to know $A^{incoh}$ to determine the conditions of
space-charge instabilities. It is surprise that we can determine
it only measuring the dc current as a function a dc bias. Using
only a finite difference we can select the working point of a
generator that is optimal for its stability. Note that
$A^{incoh}>0$ in the quasistatic case and the system is stable
against small fluctuations of internal field.

On the other hand, following Eq.~(\ref{coh}) finding of the
coherent component of absorption at the high frequency
corresponding to $m$th harmonic of the pump frequency
($\omega_2=m\omega_1$) is reduced to \textit{the calculation of
$2m$th harmonic of the current within quasistatic approach}. That
is, one needs to know
\begin{equation}
\label{harm-Vm} V_m^{harm}(U_{dc})=2\left\langle
I^{ET}(U_{dc}+U_{ac}\cos(\omega_1t))\cos(m\omega_1t)\right\rangle_t
\end{equation}
(Note that in comparison with (\ref{harm-gen}) we have in
(\ref{harm-Vm}): $I(U)\rightarrow I^{ET}(U)$). In analogy with
$A^{incoh}$, the coherent absorption $A^{coh}$ can be found if we
know only the dependence on dc bias. $A^{coh}$ is proportional to
the quantum derivative of $A^{harm}$ in the semiquasistatic case,
and $A^{coh}$ goes into a common derivative in the quasistatic
case. These derivatives are shown in the Figure~\ref{fig:2}
($\langle V_m\rangle\equiv A^{harm}(m\omega_1)$)) for the case of
7 harmonic.

To the best of our knowledge this paper is the first work, where
the quantum derivatives naturally appeared in a description of
microstructure's response to a bichromatic field. It is the
exciting problem to understand whether our semiquasistatic
approach, developed for superlattices operating in the miniband
transport regime, can be generalized to other tunnelling
structure.

\ack This research was supported by Academy of Finland (grant
109758) and AQDJJ Programm of European Science Foundation.

\end{document}